\documentclass{IEEEtran}
\usepackage{cite}
\usepackage{amsmath,amssymb,amsfonts}
\usepackage{graphicx}
\usepackage{textcomp,nicefrac}
\def\BibTeX{{\rm B\kern-.05em{\sc i\kern-.025em b}\kern-.08em
T\kern-.1667em\lower.7ex\hbox{E}\kern-.125emX}}
\usepackage[colorlinks,linkcolor=blue,anchorcolor=blue,citecolor=blue]{hyperref}
\usepackage[switch]{lineno}

\begin{document}

\title{Time Resolution Characterization of 4H-SiC LGADs with a ${}^{90}$Sr Source}
\author{Tao Yang, Yashas Satapathy, Ben J. Sekely, Abraham Tishelman-Charny, Greg Allion, Gil Atar, Philip Barletta, Carl Haber, Steve Holland, John F. Muth, Spyridon Pavlidis, and Stefania Stucci
\thanks{This work is supported by the U.S. Department of Energy, Office of Science, Office of High Energy Physics, under Contract Number DE-AC02-05CH11231 and Award Number DE-SC0024252. Additional support was performed in part at the NCSU Nanofabrication Facility (NNF), a member of the North Carolina Research Triangle Nanotechnology Network (RTNN), which is supported by the National Science Foundation (Grant ECCS-1542015) as part of the National Nanotechnology Coordinated Infrastructure (NNCI).}
\thanks{Tao Yang, Steve Holland and Carl Haber are with Physics Division, Lawrence Berkeley National Laboratory, 1 Cyclotron Road, Berkeley, 94720, CA, U.S (e-mail: chhaber@lbl.gov).}
\thanks{Ben J. Sekely, Yashas Satapathy, Greg Allion, Gil Atar, Philip Barletta, Spyridon Pavlidis and John F. Muth are with Department of Electrical and Computer Engineering, North Carolina State University, 2410 Campus Shore Dr, Raleigh, 27606, NC, U.S. (e-mail: muth@ncsu.edu).}
\thanks{Stefania Stucci and Abraham Tishelman-Charny are with Brookhaven National Laboratory, 98 Rochester St, Upton, 11973, NY, U.S.}
}

\maketitle


\begin{abstract}
This work presents timing measurements of 4H-SiC Low Gain Avalanche Detectors (4H-SiC LGADs) using beta particles from a ${}^{90}$Sr source. The 4H-SiC LGADs exhibit fast signal responses, and a time resolution of 61~ps was achieved, comparable to that of standard Si LGADs. The present limitation in the time resolution of 4H-SiC LGADs appears to stem from limited charge generation. Nevertheless, their higher voltage tolerance and faster carrier drift suggest that, with increased charge collection, their timing performance could approach or even surpass that of Si LGADs. These results demonstrate the strong potential of 4H-SiC LGADs as a robust platform for precision timing in future 4D tracking detectors, while also highlighting that signal charge is the dominant factor currently limiting their performance, indicating that further optimization of gain and drift structures will be essential for future development.
\end{abstract}


\section{Introduction} \label{sec:introduction}
As a wide bandgap semiconductor, 4H-SiC has attracted increasing attention in recent years for its potential application in collider detectors \cite{EFCA_Road_Map_2021,Basic_Research_Needs_2019}, due to its low leakage current, fast carrier saturation drift velocity, and excellent thermal stability \cite{SiC_Rad_Detecor_Review}. However, the relatively small signal charge generated by minimum ionizing particles (MIPs) (only about 2/3 that of silicon \cite{SiC_MIPs}), along with the technical challenges associated with growing high-quality, low-doped epitaxial SiC epitaxial layers, have limited the widespread adoption of SiC detectors in high-energy physics experiments.

Previous experimental results have shown that conventional SiC PiN devices already achieve a time resolution with a hundred picoseconds  \cite{SiC_Alpha_Time,Tao_NJU_PIN_Time,SiC_Rad_Detecor_Review}. However, the relatively small charge signal imposes significant challenges in achieving fast single-MIP detection with time resolution better than 100~ps for SiC devices. Inspired by the success of Si LGADs \cite{CNM_LGAD,HPK_LGAD,FBK_LGAD,BNL_LGAD,KeweiWU_LGAD,YunyunFAN_NDL}, a promising solution to these limitations is the development of 4H-SiC Low Gain Avalanche Detectors (4H-SiC LGADs) \cite{Carl_SiC_LGAD, Tao_Design_SiC_LGAD}. 4H-SiC LGADs not only provide enhanced signal charges through internal gain, but also inherit the intrinsic material advantages of SiC and the excellent timing performance of the LGAD architecture, making them highly suitable for applications in 4D tracking. Recently, 4H-SiC LGADs have been successfully fabricated \cite{Tao_Characterization_SiC_LGAD}, and their excellent timing performance with better than 35~ps time resolution has been demonstrated using the ultraviolet transient current technique (UV-TCT) \cite{Tao_EDL}. Building upon these previous results, this work characterizes the time resolution of 4H-SiC LGADs using beta particles from a ${}^{90}\mathrm{Sr}$ source, demonstrating its potential for effective use in timing applications with MIPs.

\section{Device Characteristics} \label{sec:device_structure}

\begin{figure}[!t]
\centerline{\includegraphics[width=0.98\columnwidth]{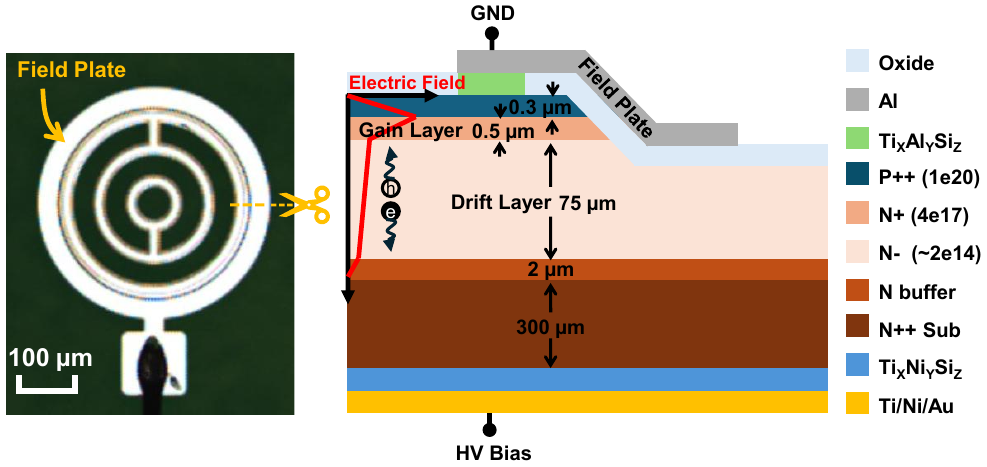}}
\caption{Photograph of the 300~${\mu}m$ thick 4H-SiC LGAD (left) and diagram of the cross-sectional view of the device (right).}
\label{fig:device_structure}
\end{figure}

\begin{figure}[!t]
\centerline{\includegraphics[width=0.98\columnwidth]{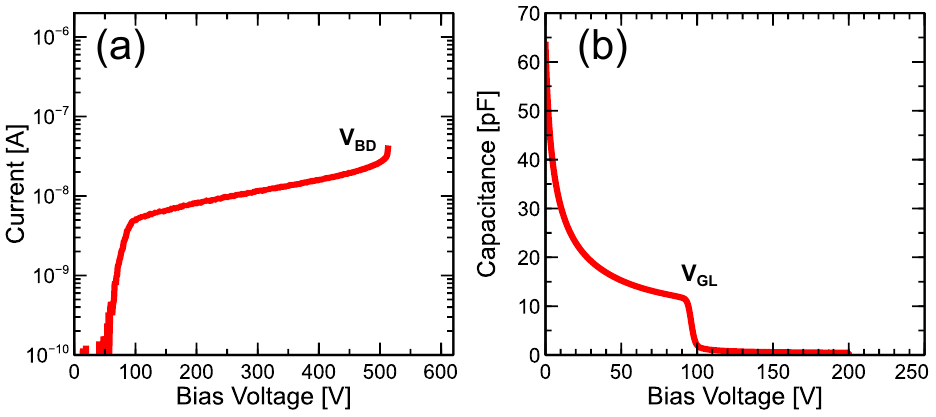}}
\caption{(a) I–V and (b)~C–V characteristics of the 300~$\mu m$ 4H-SiC LGAD.}
\label{fig:iv_cv}
\end{figure}

A single 4H-SiC LGAD used for this study and a schematic illustration of its cross-sectional structure are shown in Fig.~\ref{fig:device_structure}. The device has a diameter of 300~${\mu}m$ and incorporates a field plate structure designed to effectively enhance the breakdown voltage \cite{Tao_EDL}. The epitaxial structure consists of a $(P^{++})$–$(N^{+})$–$(N^{-})$ layer sequence: a 0.3~${\mu}m$ thick $P^{++}$ contact layer with a doping concentration of $1\times10^{20}~cm^{-3}$, a 0.5~${\mu}m$ thick $N^{+}$ gain layer with a doping concentration of $4\times10^{17}~cm^{-3}$, and a 75~${\mu}m$ thick $N^{-}$ drift layer with a doping concentration of $\sim2\times10^{14}~cm^{-3}$. When a reverse bias of 500~V is applied, the gain layer forms a wedge-shaped electric field exceeding 3~MV/cm, enabling low-gain charge multiplication as carriers drift into this high-field region. This 4H-SiC LGAD achieves a gain of 7–8 which is measured by UV-TCT \cite{Tao_EDL}.  Fig.~\ref{fig:iv_cv}(a) shows the I–V characteristic of the 4H-SiC LGAD, indicating a breakdown voltage $V_{BD}$ of 520~V and a leakage current below 10~nA prior to breakdown. The C–V curve in Fig.~\ref{fig:iv_cv}(b) exhibits a characteristic capacitance drop due to the multilayer doping structure of the LGAD, with the depletion voltage of the gain layer $V_{\mathrm{GL}}$ identified at approximately 100~V. The capacitance then drops sharply, and with further increase in voltage, the drift layer begins to deplete. With the increase in depletion region thickness, the device collects more charge generated by MIP ionization, which then drifts toward the gain layer and undergoes avalanche multiplication to produce additional charge carriers.

\section{Test Setup} \label{sec:test_setup}

\begin{figure}[!t]
\centerline{\includegraphics[width=0.98\columnwidth]{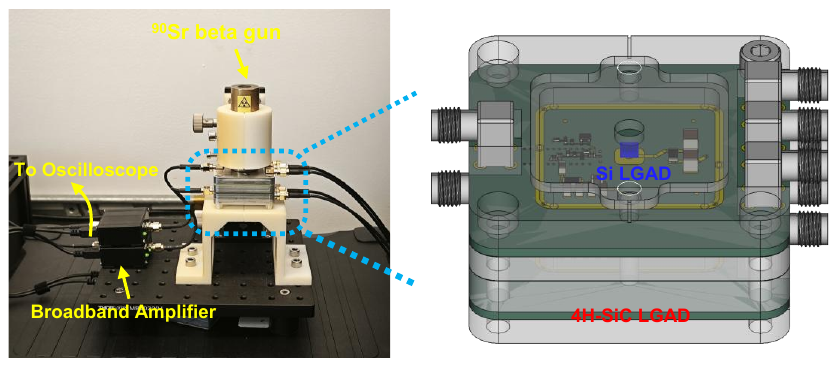}}
\caption{Photograph of the ${}^{90}\mathrm{Sr}$ source test setup. The Si LGAD was mounted on the upper circuit board, and the 4H-SiC LGAD was placed on the lower circuit board.}
\label{fig:onsite_setup}
\end{figure}

\begin{figure}[!t]
\centerline{\includegraphics[width=0.98\columnwidth]{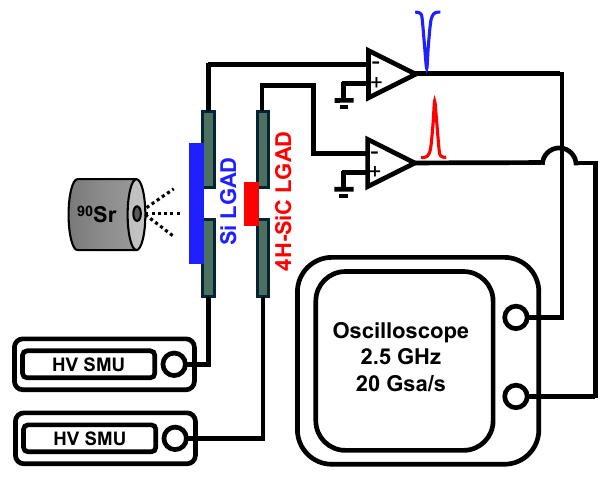}}
\caption{Schematic of the ${}^{90}\mathrm{Sr}$ source test setup.}
\label{fig:setup_scheme}
\end{figure}

\begin{figure}[!t]
\centerline{\includegraphics[width=0.9\columnwidth]{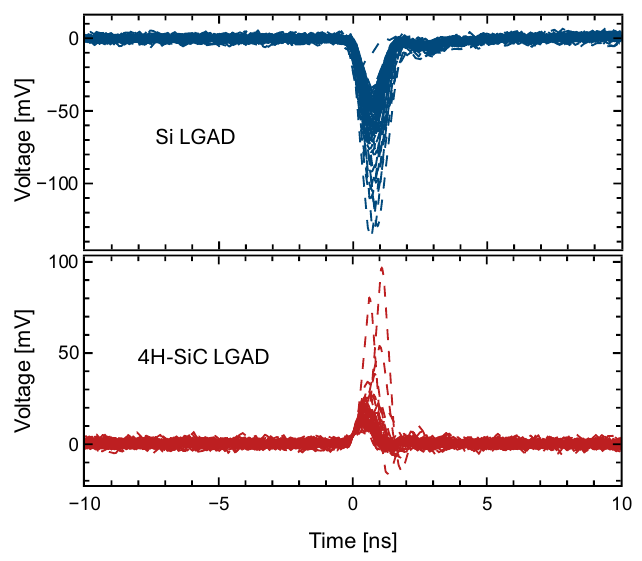}}
\caption{Sampled waveforms with the ${}^{90}\mathrm{Sr}$ source: (a)~pulse signal detected by the Si LGAD; (b)~pulse signal detected by the 4H-SiC LGAD.}
\label{fig:waveform}
\end{figure}

To measure the time resolution of the 4H-SiC LGAD, a 50~${\mu}m$ thick Si LGAD with a gain of 10-20 from Brookhaven National Laboratory (BNL) \cite{BNL_LGAD} was used as the timing reference device in the ${}^{90}\mathrm{Sr}$ source setup. The time resolution of the BNL Si LGAD has been previously determined using a ${}^{90}\mathrm{Sr}$ source with a value about 41~ps. Fig.~\ref{fig:onsite_setup} shows the ${}^{90}\mathrm{Sr}$ source test system. As illustrated in Fig.~\ref{fig:setup_scheme}, two high-speed readout boards, each equipped with a 2~GHz bandwidth transimpedance amplifier (TIA), were stacked together. The front board was mounted with a Si LGAD, while the rear board housed the 4H-SiC LGAD. To minimize scattering and attenuation of the incident beta particles by the PCB, each board featured a 2 mm diameter through-hole at the center of the device mounting pad. The ${}^{90}\mathrm{Sr}$ source emits a continuous beta spectrum with a maximum energy of 0.546~MeV; a portion of the particles can penetrate both the Si LGAD and the 4H-SiC LGAD, inducing current pulses in each detector. These signals were amplified by the transimpedance amplifier followed by secondary broadband amplifiers, and then captured by a high-speed oscilloscope. Since both the TIA on the readout board and the external wide-band amplifier are inverted, the polarity of the final output pulse signal remains unchanged.

\begin{figure}[!t]
\centerline{\includegraphics[width=0.98\columnwidth]{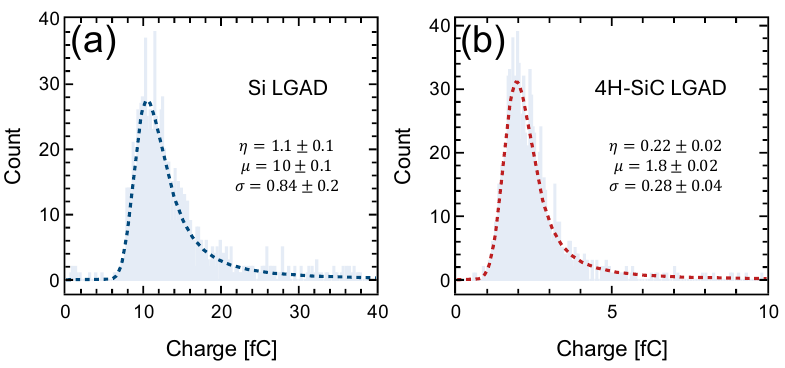}}
\caption{Charge collection distribution of the ${}^{90}\mathrm{Sr}$ source and Landau-Gaussian convolution fit results: (a)~Si LGAD at 250~V; (b)~4H-SiC LGAD at 500~V.}
\label{fig:collected_charges}
\end{figure}

\begin{figure}[!t]
\centerline{\includegraphics[width=0.98\columnwidth]{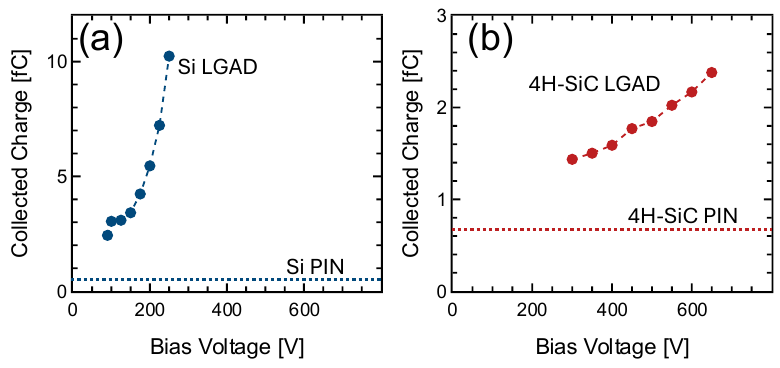}}
\caption{(a)~Collected charge of the Si LGAD as a function of bias voltage, with the dotted line indicating the theoretical collected charge for MIPs in a Si PiN detector of the same thickness;
(b)~Collected charge of the 4H-SiC LGAD as a function of bias voltage, with the dotted line indicating the theoretical collected charge for MIPs in a 4H-SiC PiN detector of the same thickness.}
\label{fig:charge_voltage}
\end{figure}


\section{Timing Performance} \label{sec:timing}

\begin{figure}[!t]
\centerline{\includegraphics[width=0.98\columnwidth]{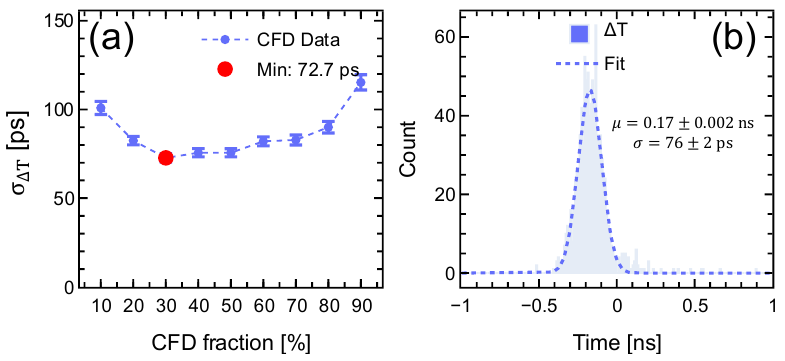}}
\caption{4H-SiC LGAD at a bias voltage of 500 V: (a) ${\sigma}_{\Delta T}$ obtained using a 10\%–90\% fraction; (b) ${\Delta T}$ distribution at a 40\% fraction with Gaussian fit.}
\label{fig:cfd}
\end{figure}

To measure the time resolution, the same method described in \cite{Tao_NJU_PIN_Time} was employed. Using the constant fraction discrimination (CFD) technique, the timing points at the same fraction of the pulse amplitude were extracted for both the Si LGAD and the 4H-SiC LGAD, denoted as $t_{\mathrm{Si}}$ and $t_{\mathrm{SiC}}$, respectively. The time difference was then calculated as ${\Delta}T = t_{\mathrm{SiC}} - t_{\mathrm{Si}}$. The time spread ${\sigma}_{\Delta T}$ was extracted by performing a Gaussian fit to the distribution of ${\Delta}T$. As shown in Fig.~\ref{fig:cfd}(a), ${\sigma}_{\Delta T}$ reaches its minimum for fractions between 30\% and 50\%. Fig.~\ref{fig:cfd}(b) shows the ${\Delta T}$ distribution and fitting result for a 40\% fraction. In this work, a 40\% fraction was consistently used to achieve the optimal time resolution. Since ${\sigma}^2_{\Delta T} = {\sigma}^2_{\mathrm{Si}} + {\sigma}^2_{\mathrm{SiC}}$, and the time resolution of the Si LGAD  ${\sigma}_{\mathrm{Si}}$, is known, the time resolution of the 4H-SiC LGAD ${\sigma}_{\mathrm{SiC}}$ can be extracted by deconvolving ${\sigma}_{\mathrm{Si}}$ from ${\sigma}_{\Delta T}$ in quadrature.

\begin{figure}[!t]
\centerline{\includegraphics[width=0.9\columnwidth]{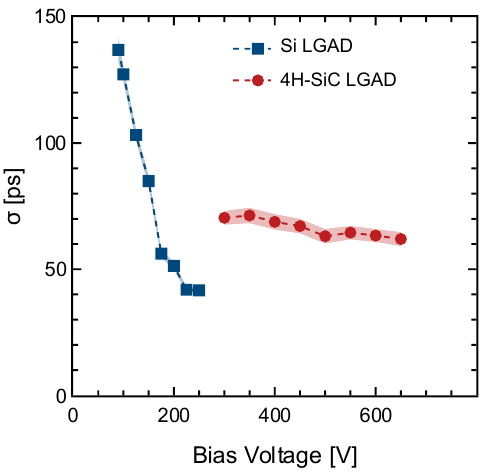}}
\caption{Time resolution of the Si LGAD and 4H-SiC LGAD as a function of bias voltage.}
\label{fig:time_voltage}
\end{figure}

Fig.~\ref{fig:time_voltage} shows the time resolution results measured by the ${}^{90}\mathrm{Sr}$ source. The 4H-SiC LGAD achieves a time resolution of 61~ps when measured with beta particles from the ${}^{90}\mathrm{Sr}$ source, which is comparable to that of conventional Si LGADs with $\sim$50~ps \cite{Si_LGAD_2022} and better than that of 4H-SiC PiN with $\sim$100~ps \cite{SiC_Alpha_Time,Tao_NJU_PIN_Time}. It is also observed that the time resolution of the 4H-SiC LGAD measured with beta particles is worse than that of the Si LGAD. It is likely due to the smaller charge generated by beta particles in the 4H-SiC LGAD. However, the faster carrier drift velocity at high bias voltage enables the device to still reach a time resolution of 61~ps.

\begin{figure}[!t]
\centerline{\includegraphics[width=0.9\columnwidth]{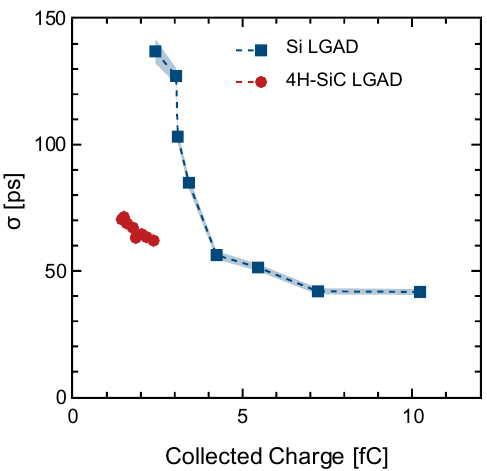}}
\caption{Time resolution of the Si LGAD and 4H-SiC LGAD as a function of collected charges}
\label{fig:time_charge}
\end{figure}

Fig.~\ref{fig:time_charge} shows the measured time resolution as a function of collected charge. It confirms that beta particles produce significantly more charge in the Si LGAD than in the 4H-SiC LGAD, consistent with the charge distribution results shown in Fig.~\ref{fig:collected_charges} and Fig.~\ref{fig:charge_voltage}. These results indicate that the present limitation in the time resolution of 4H-SiC LGADs may be associated with the limited charge generation. Although the 4H-SiC LGAD studied in this work has a modest internal gain, the relatively high drift layer doping concentration and breakdown voltage hinder full depletion of the 4H-SiC LGAD under the available bias voltage, restricting charge generation. In addition, the intrinsically smaller charge yield of beta particles in 4H-SiC due to due to the lower ionization energy deposition further reduces the collected charge. Therefore, optimizing the device gain, as well as the thickness and doping concentration of the drift layer to increase charge collection, will be essential for future development of 4H-SiC LGADs. However, Fig.~\ref{fig:time_charge} also indicates that, for the 4H-SiC LGAD, even when the collected charge is small, the timing performance can be improved by increasing the bias voltage to enhance carrier drift velocity. Under the same collected charge, 4H-SiC LGADs benefit from higher voltage tolerance and faster carrier drift compared to Si LGADs. With higher charge collection, it is expected that the timing resolution could be further improved, potentially reaching or even surpassing that of Si LGADs.

\section{Conclusion}
In this work, the timing performance of recently developed 4H-SiC LGADs was evaluated using ${}^{90}\mathrm{Sr}$ beta particles. The fabricated 4H-SiC LGADs exhibit fast response characteristics. Despite the relatively small collected charge under beta irradiation, time resolutions of 61~ps were achieved, demonstrating the potential of 4H-SiC LGADs for precision timing applications in high energy physics experiments.

Comparative studies with standard Si LGADs indicate that the time resolution of the 4H-SiC LGADs is currently limited by the reduced signal charge due to the lower ionization energy deposition and small internal gain and partially depleted drift layer. Charge-dependent timing analysis further indicates that improving the collected charge is critical for enhancing time resolution. These results highlight the importance of further device optimization—particularly in terms of gain layer engineering and drift layer design—to maximize charge multiplication and collection.

Overall, this study provides the first validation of 4H-SiC LGAD with good timing performance for MIPs detection, and lays the groundwork for potential applications in high-time-resolution detectors for 4D tracking in extreme environments such as next-generation colliders.

\section*{Acknowledgment}
We gratefully acknowledge Gabriele Giacomini and Alessandro Tricoli from BNL for providing the Si LGAD samples and for their valuable assistance in device characterization.


\bibliographystyle{ieeetr}
\bibliography{p3_sic_lgad}

\end{document}